\documentclass[sort&compress,12pt]{article}%
\usepackage{amssymb}
\usepackage{amsfonts}
\usepackage{amsmath}
\usepackage{graphicx}%
\usepackage{pgfplots}%
\begin{document}

\title{Analytical Solution for Gross-Pitaevskii Equation in Phase Space and Wigner Function}
\author{A. X. Martins$^{b}$, R.A.S. Paiva$^{b}$, G. Petronilo$^{b}$,\\ R. R. Luz$^{b}$,, S.C. Ulhoa$^{b}$, R.G.G. Amorim$^{a,b}$ T.M.R. Filho\\${}^{a}$ Universidade de Bras\'{\i}lia, Faculdade Gama,\\72444-240, Bras\'{\i}lia, DF, Brazil\\${}^{b}$International Centre for Condensed Matter Physics$,$\\Instituto de F\'{\i}sica, Universidade de Bras\'{\i}lia,\\70910-900, Bras\'{\i}lia, DF, Brazil}
\maketitle

\begin{abstract}
In this work we study symplectic unitary representations for the Galilei
group. As a consequence a Non-Linear Schr\"odinger
equation is derived in phase space. The formalism is based on
the non-commutative structure of the star-product, and using the
group theory approach as a guide a physically consistent theory is constructed in
phase space. The state is described
by a quasi-probability amplitude that is in association with the
Wigner function. With these results, we solve the Gross-Pitaevskii equation in phase space and obtained the Wigner function for the system considered.

\end{abstract}

\section{Introduction}

 A relevant equation, that describes a variety of physical phenomena, as a Bose-Einstein condensed,  is the Gross-Pitaeviskii equation \cite{carr}. The Gross-Pitaevskii model is a extension of Schr\"odinger equation and is given by \cite{kasp, and}
\begin{equation}\nonumber
i\frac{\partial\psi(r,t)}{\partial t}=\left[-\frac{1}{2M}\nabla^{2}+g|\psi(r,t)|^{2}\right]\psi(r,t),
\end{equation}
where $m$ represents mass, $V_{ext}$ is interaction potential and $g$ is intensity of atomic interaction. The study of solutions for this equation is relevant in both ways, a theoretical and applied viewpoints. In addition, an important case for the Gross-Pitaevskii system is their approach in quantum phase space, particularly the calculation of Wigner function for this system, in which is non knowledge in the literature.   

In this context, the first formalism to quantum mechanics in phase space was introduced by Wigner in 1932 \cite{wig1}. He was motivated by the problem of finding a way to improve the quantum statistical mechanics. Wigner introduced his formalism by using a kind of Fourier transform of the density matrix, $\rho(q,q^{\prime}),$
giving rise to what is called nowadays the Wigner function, $f_{W}(q,p),$
where $(q,p)$ are the coordinates of a phase space manifold ($\Gamma
$)~\cite{wig1,wig2,wig3,wig4}. The Wigner function has the same content of usual wave function obtained by Schr\"odinger equation. However, Wigner function is identified as a quasi-probability density in the sense that $f_{W}(q,p)$ is real but not
positive definite, and as such cannot be interpreted as a probability.
However, the integrals $\rho(q)=\int f_{W}(q,p)\,dp$ and $\rho(p)=\int
f_{W}(q,p)\,dq$ are (true) distribution functions. The calculation of Wigner function is based in the following steps:(i) firstly, the Schr\"odinger equation for a specific potential must be solved; (ii) in sequence, using the solutions founded, the matrix density elements are calculated; (iii) finally, a kind of Fourier transform of matrix elements must be performed. We notice that in the Wigner approach is complicated to treat non-linear potentials such as Gross-Pitaevskii system. For this reason, another methods to quantum mechanics in phase space are developed in literature. An alternative method is based in the following property of Wigner formalism: in the Wigner function approach, each operator, $A$, defined in the Hilbert space, $\mathcal{H}$, is associated with a function, $a_{W}(q,p)$, in $\Gamma
$. This procedure is precisely specified by a mapping $\Omega_{W}:A\rightarrow
a_{W}(q,p)$, such that, the associative algebra of operators defined in $\mathcal{H}$ turns out to be an algebra in $\Gamma,$ given by $\Omega
_{W}:AB\rightarrow a_{W}\star b_{W},$ where the star-product, $\star$,$\,$\ is
defined by
\begin{equation}
a_{W}\star b_{W}=a_{W}(q,p)\exp\left[  \frac{i\hbar}{2}(\frac{\overleftarrow
{\partial}}{\partial q}\frac{\overrightarrow{\partial}}{\partial p}%
-\frac{\overleftarrow{\partial}}{\partial p}\frac{\overrightarrow{\partial}%
}{\partial q})\right]  b_{W}(q,p). \label{mar261}%
\end{equation}
The result is a non-commutative structure in
$\Gamma$, that has been explored in different ways~\cite{wig2}-\cite{seb13}.

Using star-operators defined in Wigner formalism, unitary representations of
symmetry Lie groups have been developed on a symplectic manifold \cite{seb1,seb2,seb22,sig1,seb222}. The unitary representation of Galilei group leads to Schr\"odinger equation in phase space. In the analog procedure the scalar Lorentz group for spin 0 and spin 1/2, leads to the Klein-Gordon and Dirac equations in phase space. In both cases, relativistic and non-relativistic, the wave functions
are closely associated with the Wigner function~\cite{seb1,seb2}. In terms of non-relativistic quantum mechanics, the proposed formalism has been used to treat a non-linear oscillator perturbatively, to study the notion of coherent states and to introduce a non-linear Schr\"{o}dinger equation from the point of view of phase-space. In the present work we apply this symplectic formalism to find
the Wigner function for the Gross-Pitaevskii model.

The paper is organized as follows.  In the section 2, we write the Non-linear equation in phase space and we present the relation between phase space amplitude an Wigner function. In section 3 we solve the Gross-Pitaevskii equation in phase space and calculate the Wigner function. In section 4 we plot graphs of Wigner function and calculate parameter of negativity associated with the system. Finally, some closing comments are given in Section 5.

\section{Non-Linear Schrödinger Equation in Phase Space}

Using the star-operators, $\widehat{A}=a(q,p)\star$, we define the momentum and position operators, respectively by

\begin{equation}
\widehat{Q}=q\star =q+\frac{i\hbar }{2}\partial _{p},  \label{eq 7}
\end{equation}

\begin{equation}
\widehat{P}=p\star =p-\frac{i\hbar }{2}\partial _{q}.  \label{eq 8}
\end{equation}%

Then, we introduce the following operators
\begin{equation}
\widehat{K}=M\widehat{Q}_{i}-t\widehat{P}_{i},  \label{eq 9}
\end{equation}
\begin{small}
\begin{eqnarray}
\widehat{L}_{i} &=&\epsilon _{ijk}\widehat{Q}_{j}\widehat{P}_{k}  \notag \\
&=&\epsilon _{ijk}q_{j}p_{k}-\frac{i\hbar }{2}\epsilon _{ijk}q_{j}\frac{%
\partial }{\partial p_{k}}+\frac{i\hbar }{2}\epsilon _{ijk}p_{k}\frac{%
\partial }{\partial q_{j}}+\frac{\hbar ^{2}}{4}\frac{\partial ^{2}}{\partial
q_{j}\partial p_{k}}\nonumber\\
\label{mar263}
\end{eqnarray}%
\end{small}
and
\begin{small}
\begin{eqnarray}
\widehat{H} &=&\frac{\widehat{P}^{2}}{2M}=\frac{1}{2M}(\widehat{P}_{1}^{2}+%
\widehat{P}_{2}^{2}+\widehat{P}_{3}^{2})  \notag  \label{eq 11} \\
&=&\frac{1}{2M}[(p_{1}-\frac{i\hbar }{2}\frac{\partial }{\partial q_{1}}%
)^{2}+(p_{2}-\frac{i\hbar }{2}\frac{\partial }{\partial q_{2}})^{2}+(p_{3}-%
\frac{i\hbar }{2}\frac{\partial }{\partial q_{3}})^{2}]\,.\nonumber\\
\end{eqnarray}%
\end{small}
These operators, given in Eq.(\ref{eq 7}-\ref{mar263}) are defined in the Hilbert space, $\mathcal{H}(\Gamma)$, constructed with complex functions in the phase space \cite{seb1}, and satisfy the set of commutation relations for the Galilei-Lie algebra, that is
\begin{equation*}
\lbrack \widehat{L}_{i},\widehat{L}_{j}]=i\hbar \epsilon _{ijk}\widehat{L}%
_{k},
\end{equation*}%
\begin{equation*}
\lbrack \widehat{L}_{i},\widehat{K}_{j}]=i\hbar \epsilon _{ijk}\widehat{K}%
_{k},
\end{equation*}%
\begin{equation*}
\lbrack \widehat{L}_{i},\widehat{P}_{j}]=i\hbar \epsilon _{ijk}\widehat{P}%
_{k},
\end{equation*}%
\begin{equation*}
\lbrack \widehat{K}_{i},\widehat{P}_{j}]=i\hbar M\delta _{ij}\mathbf{1},
\end{equation*}%
\begin{equation*}
\lbrack \widehat{K}_{i},\widehat{H}]=i\hbar \widehat{P}_{i},
\end{equation*}%
with all other commutation relations being null. This is the Galilei-Lie algebra with a central extension characterized by $M$. The operators defining the Galilei symmetry $\widehat{P}$, $\widehat{K}$%
, $\widehat{L}$ and $\widehat{H}$ are then generators of translations,
boost, rotations and time translations, respectively.
$\widehat{Q}$ and $\widehat{P}$ can be taken to be the physical observable of position and momentum. To be consistent,
generators $\widehat{L}$ are interpreted as the angular momentum operator,
and $\widehat{H}$ is taken as the Hamiltonian operator. The Casimir
invariants of the Lie algebra are given by
\begin{equation*}
I_{1}=\widehat{H}-\frac{\widehat{P}^{2}}{2m}\quad \mathrm{and}\quad I_{2}=%
\widehat{L}-\frac{1}{m}\widehat{K}\times \widehat{P},
\end{equation*}%
where $I_{1}$ describes the Hamiltonian of a free particle and $I_{2}$ is
associated with the spin degrees of freedom. First, we study the scalar
representation; i.e. spin zero.

Using the time-translation generator, $\widehat{H}$, we derive the time-evolution equation for $\psi (q,p,t)$, i.e.,

\begin{equation}
i\hbar \partial _{t}\psi (q,p;t)=H(q,p)\star \psi (q,p;t),  \label{eq 13}
\end{equation}%
which is the Schr\"odinger equation in phase space~\cite{seb1}. The function $\psi(q,p,t)$ is defined in a Hilbert space $\mathcal{H}(\Gamma)$ associated to phase space $\Gamma$ \cite{seb1}.

The association of $\psi (q,p,t)$ with the Wigner function is given by \cite%
{seb1},
\begin{equation}
f_{W}(q,p)=\psi (q,p,t)\star \psi ^{\dagger }(q,p,t).  \label{eq14}
\end{equation}%
This function satisfies the Liouville-von Neumann equation \cite%
{seb1}. This provides a complete set of physical rules to interpret representations
and opens the way to study other improvements. In this sense, the Non-Linear Schr\"odinger equation in phase space is given by
\begin{eqnarray}\label{nls1}
i\frac{\partial}{\partial t}\psi(q,p,t)&=&\left(\frac{p^2}{2M}-\frac{\hbar^2}{8M}\frac{\partial^2}{\partial q^2}-i\hbar\frac{p}{2M}\frac{\partial}{\partial q}\right)\psi(q,p,t)\nonumber\\
&+&V\left(q+\frac{i\hbar}{2}, t\right)\psi(q,p,t)+g|\psi(q,p,t)|^2\psi(q,p,t),\nonumber
\end{eqnarray}
where $g$ is intensity of atomic interaction. This equation describes several physical systems, in particular, it is used to study the Bose-Einstein condensation. Eq.(\ref{nls1}) is derived from the Lagrangian density
\begin{eqnarray}\nonumber
\mathcal{L}&=&\frac{i\hbar}{2}(\psi^{\dagger}\partial_{t}\psi - \psi\partial_{t}\psi^{\dagger})+ \frac{i\hbar}{4M}p(\psi^{\dagger}\partial_{q}\psi - \psi\partial_{q}\psi^{\dagger})\nonumber\\
&-&\frac{p^{2}}{2M}\psi\psi^{\dagger} + V(q)\star(\psi\psi^{\dagger}) - \frac{\hbar^{2}}{8M}\partial_{q}\psi\partial_{q}\psi^{\dagger}+g\psi^4.\nonumber
\end{eqnarray}

In the next section, we solve this equation and calculate the Wigner function for this system.

\section{ Solution of Gross Pitaevskii Equation and  Wigner Function}

In this section we present a solution for Non-linear Schr\"odinger equation in phase space and the associated Wigner function. 

The Gross-Pitaeviskii equation in  phase space is written by
\begin{eqnarray}\nonumber
i\frac{\partial}{\partial t}\psi(q,p,t)&=&\left(\frac{p^2}{2M}-\frac{\hbar^2}{8M}\frac{\partial^2}{\partial q^2}-i\hbar\frac{p}{2M}\frac{\partial}{\partial q}\right)\psi(q,p,t)\nonumber\\
&+&V\left(q+\frac{i\hbar}{2}, t\right)\psi(q,p,t)+g(\psi(q,p,t)\psi^{\dagger}(q,p,t))\psi(q,p,t),\nonumber
\end{eqnarray}
In this work we addressed the stationary equation without external potential, $V_{ext}=0$. In this way, the equation above becomes
\begin{equation}\label{gpph1}
\left(\frac{p^2}{2M}-\frac{\hbar^2}{8M}\frac{\partial^2}{\partial q^2}-i\hbar\frac{p}{2M}\frac{\partial}{\partial q}\right)\psi(q,p)+g(\psi(q,p)\psi^{\dagger}(q,p))\psi(q,p)=E\psi(q,p).
\end{equation}

We now consider the solution of Eq.~\eqref{gpph1} in regions of
constant potential, which may be taken to be $V(\hat{q})=0$ without loss of generality. We note first that if $\psi(q,p)$ vanishes anywhere in an interval, as for example at the edges of the box, then $\psi(q,p)$ may be taken to be purely real throughout that
interval. This can be done only if 
$$\frac{\partial \psi(q,p)}{\partial q}\ll\frac{\partial^2\psi(q,p)}{\partial q^2}.$$
Thus we may remove the absolute value symbol in Eq.~\eqref{gpph1}. So letting $M=\hbar=1$, the equation becomes an ordinary nonlinear equation for a real function:
\begin{equation}
\frac{1}{2}\bigg(p^2-\frac{1}{4}\frac{\partial}{\partial q}\bigg)\psi(q,p)+g\psi(q,p)^3-E\psi(q,p)=0.\label{gpph2}
\end{equation}
Letting $g>0$ the solution of Eq.~\eqref{gpph2} is
\begin{equation}
\psi(q,p)=A\,\text{sn}\bigg(\sqrt{k^2-p^2}\,q+\delta\bigg|m\bigg),\label{sol1}
\end{equation}
with $k>p$, and $\text{sn}(x|m)$ is the Jacobian elliptic function, $k$ and $\delta$ will be determined by the boundary conditions, while $A$ and $m$ will be determined by substituting in \eqref{sol1} into \eqref{gpph2} and normalization.
The boundary conditions
\begin{eqnarray}
\psi(0,p)=\psi(L,p)=0.
\end{eqnarray}
The boundary condition at the origin can be satisfied by taking $\delta=0$. The function $sn(x|m)$ is periodic in $x$ with period of $4K(m)$, where $K(m)$ is the elliptic integral of the first kind. Thus the boundary equations are satisfied if $k=2nK(m)/L$, where $n=1,2,3...$. The number of nodes in the solution $nth$ is $n-1$. We then solve Eq.~\eqref{gpph2} substituting \eqref{sol1} and using the Jacobian elliptic identities, this results in equation for the amplitude A and energy E,
\begin{equation}
A^2=\frac{2m[2nK(m)]^2}{L^2}\label{amp}
\end{equation}
\begin{eqnarray}
E=\frac{[2nK(m)]^2}{2L^2}(m+1).\label{energy}
\end{eqnarray}
 Eq.(\ref{sol1}) becomes
 \bigskip
\begin{equation}\label{sol2}
\psi(q,p)=\frac{\sqrt{2}\sqrt{2m}(2nK(m))}{L}\text{sn}\bigg(\sqrt{(2nK(m))^2-p^2}\,q\bigg|m\bigg),
\end{equation}
\bigskip

The wave-function and the energy are determined up to a factor m. This result is similar to what is obtained in configuration space \cite{carr}.
Using Eq.(\ref{sol2}), we calculate Wigner function for the Gross-Pitaevskii system by Eq.(\ref{eq14}). As the wave function is real, so $fw=\psi(q,p)\star\psi(q,p)$, but by \cite{wig4} $f_w\star f_w\propto f_w$ and the time independent $f_w$ obeys the two sided energy $\star$-genvalue equation
$$
H\star f_w=Ef_w,\; f_w\star H=Ef_w,
$$
we can conclud that, in the case of this solution $\psi(q,p)=f_w$.
\section{Analysis of solution}

In this section we plot the graphics to Wigner function founded in section above and calculate the negativity parameter for this system.
\subsection{Particle in a box limit}
Both the zero density linear limit and the highly  excited-state limit, gives the particle in a box limit type solution. Mathematically $m\rightarrow 0^+$ and $\text{sn}\rightarrow\sin$. Physically, $n\gg 1$. In this limit $K(m)\rightarrow\pi\bigg[1/2+m/8+O(m^2)\bigg]$ and $m\rightarrow 1/n^2\pi^2$, so the Eq.~\eqref{energy}, becomes
\begin{eqnarray}
    E&=&\frac{n^2\pi^2}{2L^2}\bigg(1+\frac{3m}{2}+O(m^2)\bigg)
\end{eqnarray}
which converges to the linear quantum mechanics particle in a box as $m\rightarrow0^+$. One may obtain this results from first order perturbation theory~\cite{sig1}.
The behavior of Wigner function given in Eq.(\ref{sol2}) for the three first energy levels can be visualized in Figures (1)-(3). 
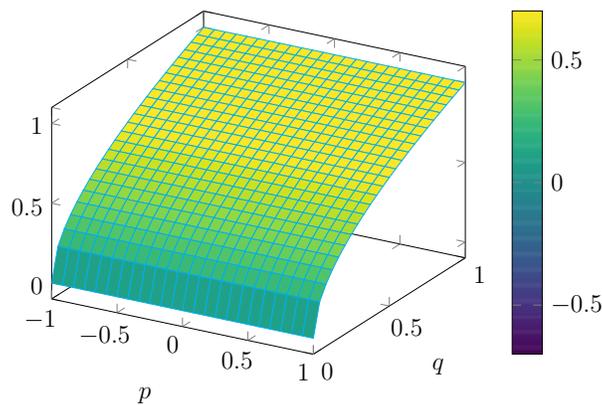
\begin{figure}[h]
\centering
\begin{tikzpicture}[scale=0.8]
\begin{axis}[point meta min=-0.7, point meta max=0.7, colorbar, colormap/viridis,
    xlabel = $p$,
    ylabel = $q$,
    view={30}{30}
]
\addplot3[
    surf, faceted
color=cyan, domain=-1:1,y domain=0:1
]
{(1/22.613)*sqrt((512-2*x^2)*y))};
\end{axis}
\end{tikzpicture}
\caption{{\protect\small {Wigner Function for Gross-Pitaevskii Model, n=100}}}
\label{figura 1}
\end{figure}

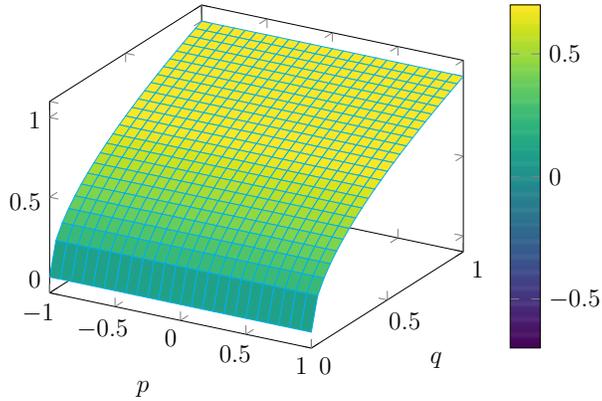
\begin{figure}[h]
\centering
\begin{tikzpicture}[scale=0.8]
\begin{axis}[point meta min=-0.7, point meta max=0.7, colorbar, colormap/viridis,
    xlabel = $p$,
    ylabel = $q$,
    view={30}{30}
]
\addplot3[
    surf, faceted
color=cyan, domain=-1:1,y domain=0:1
]
{{(1/31.99)*sqrt((1024-2*x^2)*y))}};
\end{axis}
\end{tikzpicture}
\caption{{\protect\small {Wigner Function for Gross-Pitaevskii Model, n=200}}}
\label{figura 2}
\end{figure}

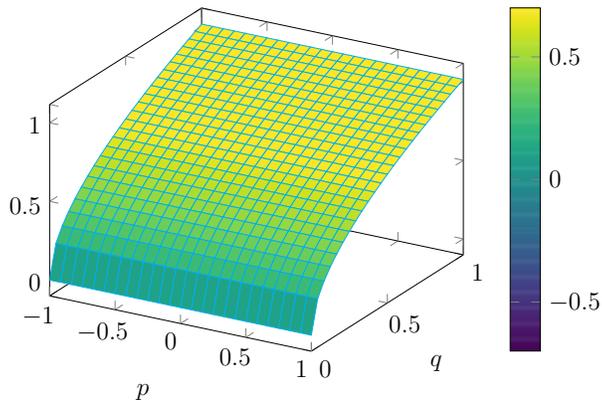
\begin{figure}[h]
\centering
\begin{tikzpicture}[scale=0.8]
\begin{axis}[point meta min=-0.7, point meta max=0.7, colorbar, colormap/viridis,
    xlabel = $p$,
    ylabel = $q$,
    view={30}{30}
]
\addplot3[
    surf, faceted
color=cyan, domain=-1:1,y domain=0:1
]
{{(1/39.183)*sqrt((1583-2*x^2)*y))}};
\end{axis}
\end{tikzpicture}
\caption{{\protect\small {Wigner Function for Gross-Pitaevskii Model, n=300}}}
\label{figura 3}
\end{figure}
\pagebreak
\newpage

Using the Wigner function, the negative parameter for the system is calculated. The results are presented in Table \eqref{t1}
\begin{table}[!htbp]\label{t1}
\caption{Parameter of negativity for $n=100,200,300$.}
\label{neg}
\scalebox{1.4}{%
 \begin{tabular}{||c c c c c c c c c ||} 
 \hline
$n$&\qquad&\qquad&\qquad\qquad&\qquad\qquad&\qquad\qquad&\qquad\qquad&\qquad&$\eta(\psi)$\\ [1ex] 
 \hline
 100 &\qquad&\qquad&\qquad\qquad&\qquad\qquad&\qquad\qquad&\qquad\qquad&\qquad\qquad& 0\\
 200&\qquad&\qquad&\qquad\qquad&\qquad\qquad&\qquad\qquad&\qquad\qquad&\qquad\qquad& 0 \\
 300&\qquad&\qquad&\qquad\qquad&\qquad\qquad&\qquad\qquad&\qquad\qquad&\qquad\qquad& 0 \\
 \hline
\end{tabular}
}
\end{table}
\pagebreak
\newpage
as the negativity parameter is correlated with the non-classicality of the system, it appears shows that in the limit where $n>>1$ the BCE behaviors is in accordance with classical mechanics. A consistent result.

\section{Concluding remarks}
In this work we studied the non-trivial problem of Gross-Pitaevskii equation in phase space, in which is a case of nonlinear Schr\"odinger. The Wigner function for this system were obtained. In our knowledge this is the first time that an analytical solution of the Wigner function for the Gross-Pitaevskii system appears in the literature. The particle in a box solution is studied for the physical meaningful limit of the solution, $n\gg 1$. We studied the parameter of negativity of the system and concluded that in these limits it appear to have a purely classical behavior. The perspective of this work is to continue studying Gross-Pitaeviskii equation in phase space for others boundary conditions and external potential as the particle in a ring and the harmonic oscillator.
\section*{Acknowledgements}

This work was partially supported by CAPES and CNPq of Brazil.

\end{document}